\documentclass[aps,prd,twocolumn,showpacs,amsmath,nofootinbib]{revtex4-1}

\usepackage{color}
\newcommand{\beqa}{\begin{eqnarray}}
\newcommand{\eeqa}{\end{eqnarray}}
\usepackage{float}
\usepackage{epsfig}
\usepackage{graphicx}
\usepackage{latexsym}
\usepackage{amsmath}
\usepackage{mathrsfs}
\usepackage{dcolumn}
\usepackage{bm}%

\begin{document}

\title{Constraints on the primordial curvature power spectrum at small scales between 
$3\times 10^{18}$ and $4.5\times 10^{21}~\rm Mpc^{-1}$}

\author{Yupeng Yang}\email{ypyang@aliyun.com}

\affiliation{School of Physics and Physical Engineering, Qufu Normal University, Qufu, Shandong, 273165, China}

\begin{abstract}
The primordial curvature power spectrum $\mathcal{P}_\mathcal{R}$ has been measured with high precision on large scales 
$10^{-4}\lesssim k\lesssim 3~\rm Mpc^{-1}$ based on observations of the cosmic microwave background, Lyman-$\alpha$ forest and 
large scale structure. On small scales $3\lesssim k \lesssim 10^{23}~\rm Mpc^{-1}$, constraints are primarily derived from studies 
on primordial black holes (PBHs). In particular, for very small scales $10^{17}\lesssim k\lesssim 10^{23}~{\rm Mpc^{-1}}$, 
current limits come exclusively from 
investigations of the lightest supersymmetric particles produced by PBH radiation and the stable Planck-mass relics after their evaporation. 
Recent findings also indicate that the evaporation of light PBHs ($M_{\rm PBH}\lesssim 10^{9}~\rm g$) can modify the expansion rate of 
the Universe and the baryon-to-photon ratio, thereby affecting the primordial abundance of light nuclei. Moreover, it has been proposed that 
the ``memory burden'' effect can slow down the mass loss rate of black holes, allowing light PBHs to survive until today.  
Based on recent theoretical advancements in black hole physics and existing constraints on the initial mass fraction of light PBHs with masses 
$10^{3}\lesssim M_{\rm PBH}\lesssim 2\times 10^{9}~\rm g$, and especially the recent constraints on memory-burdened PBHs, 
we derive new and tighter upper limits on $\mathcal{P}_\mathcal{R}$ on small scales 
$3\times 10^{18}\lesssim k\lesssim 4.5\times 10^{21}~\rm Mpc^{-1}$, a regime that has been underexplored in previous literature. 
These constraints are derived under the specific assumption that the memory burden effect activates after the PBH loses half of its 
initial mass and subsequently halts further evaporation, and different assumptions on the memory burden parameters would lead to modified limits.
 
\end{abstract}

\maketitle

\section{Introduction}

It is well known that the cosmic structures originate from early density perturbations with an amplitude of $\delta \rho/\rho\sim 10^{-5}$. 
The primordial curvature (density) perturbation power spectrum $\mathcal{P}_\mathcal{R}$ ($\mathcal{P}_\delta$) 
has been measured precisely on large scales, yielding $\mathcal{P}_\mathcal{R}\sim 10^{-8}$ for $10^{-4}\lesssim k\lesssim 3~\rm Mpc^{-1}$, 
based on studies of the cosmic microwave background (CMB), Lyman-$\alpha$ forest, 
and large scale structure~\cite{cmb_2,lyman,large,Ragavendra:2024yfp}. 
On small scales $3\lesssim k\lesssim 10^{23}~\rm Mpc^{-1}$, constraints on $\mathcal{P}_\mathcal{R}$ mainly 
come from studies of primordial black holes (PBHs). PBHs are thought to form in the very early universe via 
the gravitational collapse of large overdensities with an amplitude of $\delta \rho/\rho \gtrsim 0.3$, leading to the upper limits 
$\mathcal{P}_\mathcal{R}\lesssim 10^{-2}$~\cite{Josan:2009,mnras,Dalianis:2018ymb,PhysRevD.100.063521,prd-2020,mack_21,Caravano:2024tlp,Caravano:2024moy}. Alternatively, on scales $3\lesssim k\lesssim 10^7~\rm Mpc^{-1}$, 
tighter upper limits of $\mathcal{P}_\mathcal{R}$ can be obtained from studies of ultracompact minihalos (UCMHs)~\cite{Bringmann:2011ut,fangdali,scott_2015,yyp_neutrino,Josan:2010vn}. UCMHs could form via the collapse of 
early density perturbations with an amplitude in the range $10^{-3}\lesssim \delta \rho/\rho\lesssim 0.3$~\cite{0908.0735}. 
However, these constraints depend sensitively on the nature of the dark matter particle~\cite{Bringmann:2011ut,fangdali,scott_2015,yyp_neutrino,Josan:2010vn}. 

The lifetime of a black hole (BH) with mass $M_{\rm BH}$ scales as $\tau \sim 10^{64}(M_{\rm BH}/M_{\odot})^{3}~\rm yr$. 
Consequently, PBHs with mass $M_{\rm PBH}<10^{15}~\rm g$ are expected to have evaporated completely by now. 
For PBHs in the mass ranges $10^{9}\lesssim M_{\rm PBH}\lesssim 10^{13.5}~\rm g$ and $10^{13.5}\lesssim M_{\rm PBH}\lesssim 10^{15}~\rm g$, 
the energy released during their evaporation can affect big bang nucleosynthesis (BBN) and cosmic microwave background (CMB), 
respectively, allowing constraints to be placed on their initial mass fraction (and consequently on $\mathcal{P}_\mathcal{R}$) 
for the corresponding scales $10^{15} \lesssim k \lesssim 10^{18}~\mathrm{Mpc}^{-1}$~\cite{0908.0735,lz_decay}. 
PBHs with $M_{\rm PBH}<10^{9}~\rm g$ have a lifetime shorter than $\sim 0.01~\rm s$, evaporating well before BBN begins, 
and were thus traditionally considered unconstrained by BBN observations~\cite{carr}. However, a recent re-investigation 
in Ref.~\cite{Boccia:2024nly} shows that the evaporation of PBHs with masses $10^{8}\lesssim M_{\rm PBH}\lesssim 10^{9}~\rm g$ 
can still modify the expansion rate of the Universe and the baryon-to-photon ratio, thereby affecting the primordial abundances of 
light nuclei. Consequently, observations of helium ($^4{\rm He}$) and deuterium ($\rm D$) abundances can be used 
to derive upper limits on the initial mass fraction of PBHs for this mass range~\cite{Boccia:2024nly}.

While standard Hawking radiation predicts that light PBHs ($M_{\rm PBH}\lesssim 10^{15}~\mathrm{g}$) do not survive until the present, 
this calculation neglects the back reaction of the emission on the quantum state of the black hole, an effect that becomes significant when the energy of the emitted quanta is comparable to the total mass of the black hole~\cite{Preskill:1992tc,Dvali:2020wft}. Recently, it has been proposed that this back reaction leads to a “memory burden” effect, which can substantially slow down the mass loss rate of a black hole~\cite{Dvali:2018xpy,Dvali:2020wft,Dvali:2024hsb,Thoss:2024hsr}. As a result, PBHs with initial masses $M_{\rm PBH} \lesssim 10^{15}~\mathrm{g}$ could potentially survive and evaporate only at late times. For instance, a memory-burdened PBH with a current mass of $M_{\rm PBH} \sim 10^8~\mathrm{g}$ could still be emitting high-energy photons and neutrinos~\cite{Chianese:2024rsn}. This effect typically becomes important when the black hole mass has decreased to roughly half its initial value due to standard Hawking evaporation. In the following, we assume that once the memory burden dominates, the PBH’s subsequent mass loss is negligible. Under this assumption, two such memory-burdened PBHs could merge to form a new black hole, which would then emit high-energy particles. This scenario has been used in Ref.~\cite{Zantedeschi:2024ram} to derive constraints on the current abundance of PBHs in the mass range $10^{3} \lesssim M_{\rm PBH} \lesssim 2\times 10^{9}~\mathrm{g}$. 

In this work, we do not recalculate the PBH abundance constraints from observational data; instead, we directly adopt the latest upper limits on the initial mass fraction $\beta(M_{\rm PBH})$ from BBN~\cite{Boccia:2024nly} and on the current abundance 
$f_{\rm PBH}$ of memory-burdened PBHs from gamma-ray and neutrino observations~\cite{Thoss:2024hsr,Zantedeschi:2024ram,Chianese:2024rsn}. 
Our original contribution is to convert these existing constraints into upper limits on the primordial curvature power spectrum 
$\mathcal{P}_\mathcal{R}$ on the small scale range $3\times 10^{18} \lesssim k \lesssim 4.5\times 10^{21}~\mathrm{Mpc}^{-1}$, 
which has been largely unexplored in previous literature. By applying the memory-burdened PBH scenario, we obtain significantly tighter constraints than those from the traditional stable relics and supersymmetric particles.  

This paper is organized as follows. In Sec. II, we first briefly review the theoretical relation between the initial mass fraction of PBHs and the primordial power spectrum. We then summarize the recent observational constraints on the (initial) mass fraction of light PBHs. Finally, we translate these constraints into new limits on $\mathcal{P}_\mathcal{R}$. Our conclusions are presented in Sec. III.


\section{constraints on light PBHs and primordial curvature power spectrum at small scales }

\subsection{The initial mass fraction of PBHs}
\label{sec:mass_fraction}

Many mechanisms have been proposed for the formation of PBHs. The most popular one is that PBHs form via the gravitational collapse of early large density perturbations when they re-enter the particle horizon (see, e.g., Ref.~\cite{carr} and references therein). For a Gaussian distribution of the primordial density perturbations, the probability distribution of the smoothed density contrast $\delta(R)$ on a scale $R$ can be written as~\cite{carr,PhysRevD.100.063521} 

\beqa
P(\delta(R))=\frac{1}{\sqrt{2\pi}\sigma(R)}{\rm exp}\left(-\frac{\delta^{2}(R)}{2\sigma^{2}(R)}\right),
\label{eq:m_ucmh}
\eeqa 
where the mass variance $\sigma^{2}(R)$ is given by 

\beqa
\sigma^{2}(R)=\int^{\infty}_{0} W^{2}(kR)\mathcal{P}_{\delta}(k,t)\frac{dk}{k} 
\label{eq:sigma2}
\eeqa 
with $W(kR)={\rm exp}(-k^{2}R^{2}/2)$ being a Gaussian window function. Here $\mathcal{P}_{\delta}$ denotes 
the primordial density perturbation power spectrum. 

Within the Press-Schechter theory, the initial mass fraction of PBHs (the fraction of the Universe’s energy density that collapses into PBHs of mass $M_{\rm PBH}$ at formation) can be written as~\cite{carr} 

\beqa
\beta(M_{\rm PBH})=&&2\int^{1}_{\delta_c}P(\delta(R))d\delta(R)	\nonumber \\
=&&\frac{2}{\sqrt{2\pi}\sigma(R)}\int^{1}_{\delta_c}{\rm exp}\left(-\frac{\delta^{2}(R)}{2\sigma^{2}(R)}\right)d\delta(R)\nonumber \\
\simeq &&{\rm erfc}\left(\frac{\delta_c}{\sqrt{2\pi}\sigma(R)}\right),
\label{eq:betampbh}
\eeqa 
where $\delta_c$ is the critical density contrast required for PBH formation, which depends on the details of the collapse process. 
A simple analytical estimate gives $\delta_{c}\sim 0.3$ during the radiation dominated epoch~\cite{carr}. 
More refined numerical simulations have shown that $\delta_{c}$ lies in the range $0.42\lesssim \delta_{c}\lesssim 0.66$, 
depending on the curvature profile~\cite{Polnarev:2006aa}. In this work we adopt $\delta_{c}= 0.42$ for our calculations. 

The primordial density perturbation power spectrum $\mathcal{P}_{\delta}$ is related to the primordial curvature perturbation 
power spectrum $\mathcal{P}_{\mathcal{R}}$ by~\cite{Josan:2009,Bringmann:2011ut,PhysRevD.100.063521}

\beqa
\mathcal{P}_{\delta}=\frac{4(1+w)^{2}}{(5+3w)^{2}}\left(\frac{k}{aH}\right)^{4}\mathcal{P}_{\mathcal{R}},
\label{eq:p}
\eeqa
where $w$ is $1/3$ in the radiation dominated epoch. 

The initial mass fraction of PBH, $\beta(M_{\rm PBH})$, is related to the present density parameter $\Omega_{\rm PBH}$ via 
$\beta(M_{\rm PBH})=\Omega_{\rm PBH}(g^{\rm eq}_{\star}/g^{i}_{\star})^{1/12}(M_{\rm H}/M^{\rm eq}_{\rm H})^{1/2}$~\cite{Josan:2009}, 
where $g^{\rm eq}_{\star}$ and $g^{i}_{\star}$ are the total numbers of effectively massless degrees of freedom at matter-radiation equality and 
at the time of PBH formation, respectively. 
$M^{\rm eq}_{\rm H}$ denotes the horizon mass at matter-radiation equality. The PBH mass $M_{\rm PBH}$ is related to the horizon mass $M_{\rm H}$ 
by $M_{\rm PBH}=\gamma M_{\rm H}$, where $\gamma$ is the fraction of the horizon mass that collapses into a PBH.
Assuming that the comoving number density of PBHs is conserved after formation, the initial mass fraction $\beta(M_{\rm PBH})$ 
can be expressed in terms of the current abundance parameter $f_{\rm PBH} \equiv \Omega_{\rm PBH}/\Omega_{\rm DM}$ via~\cite{Josan:2009,carr}

\beqa
\beta(M_{\rm PBH})=1.6\times 10^{-25}f_{\rm PBH}\left(M_{\rm PBH}/{\rm g}\right)^{1/2},
\label{eq:initial_f}
\eeqa 
where we adopt the typical values $\gamma=0.2$ for the fraction of the horizon mass that collapses into a PBH, 
and $g^{i}_{\star} \approx 100$ and $g^{\rm eq}_{\star} \approx 3$ for the effective number of relativistic degrees of freedom at formation 
and at matter-radiation equality, respectively~\cite{carr}.

The (initial) mass fraction of PBHs has been constrained through a variety of cosmological and astrophysical observations (see, e.g., Refs.~\cite{carr,Su:2024hrp,Cang:2020aoo,Tashiro:2021xnj,Mena:2019nhm,Lu:2019ktw,Cai:2020fnq,Chen:2016pud,Inoue:2017csr,Ali-Haimoud:2016mbv,yinzhema,DeLuca:2020agl,Yang:2020zcu,Khlopov:2008qy} and references therein). In the following, we focus on recent constraints for light PBHs in the mass range $10^{3}\lesssim M_{\rm PBH}\lesssim 10^{9}~\mathrm{g}$, which have been less explored in earlier literature.

\subsection{Constraints on the (initial) mass fraction of light PBHs}

\subsubsection{New constraints from big bang nucleosynthesis} 

The effects of light PBHs with masses $10^{9} \lesssim M_{\rm PBH} \lesssim 10^{13.5}~\mathrm{g}$ on big bang nucleosynthesis (BBN) have been widely studied in previous literature, with studies focusing primarily on the impact of high‑energy particles from PBH evaporation on nuclear reaction rates~\cite{carr}. Recently, however, the authors of Ref.~\cite{Boccia:2024nly} showed that even these lighter PBHs in the mass range $10^{8} \lesssim M_{\rm PBH} \lesssim 10^{9}~\mathrm{g}$ can affect BBN through a different channel.

PBHs formed with masses $10^{8} \lesssim M_{\rm PBH} \lesssim 10^{9}~\mathrm{g}$ evaporate completely during the early radiation‑dominated era. Their evaporation products contribute to the radiation energy density, thereby modifying the Hubble rate as $H^{2} = (8\pi G/3)(\rho_r + \rho_{\rm PBH})$. This change in expansion rate alters the predicted mass fractions of light elements. Moreover, the entropy injected by the evaporating PBHs can shift the baryon‑to‑photon ratio $\eta$, which in the standard cosmology remains constant and directly influences the light‑element abundances.

Using the observed primordial abundances of helium ($^4\mathrm{He}$) and deuterium (D), Ref.~\cite{Boccia:2024nly} derived constraints on the initial mass fraction $\beta_{\rm PBH}$ for PBHs in the mass range $10^{8} \lesssim M_{\rm PBH} \lesssim 10^{9}~\mathrm{g}$, a region that had not been covered by earlier studies. The resulting bounds on $\beta_{\rm PBH}$ from BBN (based primarily on deuterium, because helium provides only a weak constraint) are shown by the green shaded area (right $y$‑axis) in Fig.~\ref{fig:cons_frac}.

\subsubsection{New constraints with `memory burden' effect} 

The standard Hawking evaporation calculation assumes a fixed background geometry and neglects the backreaction of the emitted quanta on the black hole’s quantum state. This approximation breaks down when the black hole possesses an exceptionally large information storage capacity quantified by its Bekenstein-Hawking entropy 
$S_{\rm BH}=4\pi M^{2}_{\rm BH}/M^{2}_{P}$ ($M_{\rm P}$ is the Planck mass) and when the energy of the emitted quanta becomes comparable to the total mass of the black hole. The resulting phenomenon, known as the memory burden effect~\cite{Dvali:2018xpy,Dvali:2020wft,Dvali:2024hsb,Thoss:2024hsr}, arises from the assisted gaplessness mechanism in systems with enhanced memory capacity. 

In the effective description developed, e.g., in~\cite{Dvali:2024hsb}, the black hole is modeled by two types of degrees of freedom. 
The master mode (denoted $\phi$) represents the graviton condensate that gives rise to the classical metric; 
its energy gap is $m_{\phi}\sim 1/R$ ($R$ is the radius of black hole), 
and its initial occupation number is $n_{\phi}\sim S_{\rm BH}$. The memory modes (denoted $\theta^{j}$ with $j=1,\dots,M$) 
are the gapless graviton harmonics that store quantum information, in the asymptotic vacuum their gaps are $m_{j}\sim M_{P}$, 
but inside the black hole they become gapless when the master mode is critically occupied. The coupling between the two sets is of order 
$1/N_{\phi}$ with $N_{\phi}\sim S_{\rm BH}$. 

The assisted gaplessness mechanism implies that when $n_{\phi}$ equals its critical value $N_{\phi}$, the effective gaps of the memory modes vanish: 
$\omega_{j}=(1-n_{\phi}/N_{\phi})^{q}m_{j}$, where $q\geq 2$ is a model dependent exponent. At this critical point, the system can store an enormous amount of information with minimal energy cost its microstate entropy thus saturates the area law. The key point is that deviating from this critical state 
requires overcoming an energy barrier, because the memory modes acquire positive gaps. This resistance to change is the memory burden.   

For a black hole, the initial state corresponds to $n_{\phi}=N_{\phi}$. As it loses mass via standard Hawking radiation, $n_{\phi}$ decreases. 
The memory burden becomes dominant when the induced gaps $\omega_{j}$ grow to be comparable to $m_{\phi}$. The condition for this transition is $\Delta n_{\phi}/N_{\phi}\sim (qM_{p}R)^{-1/(q-1)}$. For $q=2$, this is $\sim 1/\sqrt{S_{\rm BH}}\ll 1$, 
while for larger $q$ it approaches 1/2. Thus, regardless of $q$, the memory burden sets in at the latest when the black hole has lost about half 
of its initial mass. Once the memory burden dominates, further mass loss is severely suppressed. 
 
\begin{figure}
\centering
\includegraphics[width=0.5\textwidth]{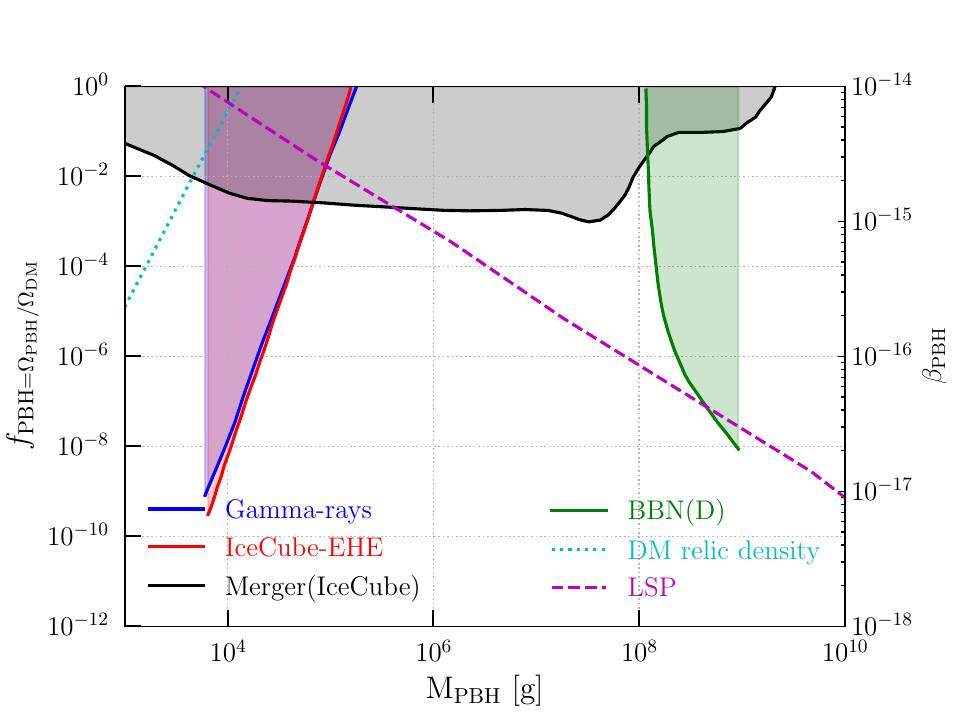}
\caption{New constraints on the (initial) mass fraction of light primordial black holes (PBHs) 
for the mass range $1\times 10^{3}\lesssim M_{\rm PBH}\lesssim 2\times10^{9}~\rm g$ are presented: 
(i) the initial mass fraction ($\beta_{\rm PBH}$) from big bang nucleosynthesis (BBN) using the deuterium abundance 
(``BBN(D)'', green shaded area, right $y$-axis)~\cite{Boccia:2024nly}; 
(ii) the present mass fraction ($f_{\rm PBH}$) of memory-burdened PBHs from gamma-ray observations (``Gamma-rays'', blue shaded area)~\cite{Thoss:2024hsr}; 
(iii) the present mass fraction ($f_{\rm PBH}$) of memory-burdened PBHs from high-energy neutrino 
observations using 7-year IceCube-EHE data (``IceCube-EHE'', red shaded area)~\cite{Chianese:2024rsn}; 
(iv) the present mass fraction ($f_{\rm PBH}$) of merged memory-burdened PBHs from high-energy neutrino observations using 
7-year IceCube data for energies above 60 TeV (``Merger (IceCube)'', black shaded area)~\cite{Zantedeschi:2024ram}. 
All memory-burdened PBH constraints adopt the exponent $k'=2$, corresponding to the most conservative estimate~\cite{Dvali:2024hsb}. 
For comparison, previous constraints on the initial PBH mass fraction ($\beta_{\rm PBH}$, right $y$-axis) 
are also plotted: a stable Planck-mass relic from PBH evaporation (``DM relic density'', dotted line) and the lightest 
supersymmetric particles (``LSP'', hashed line) produced by PBH evaporation~\cite{Josan:2009}. Both scenarios are consistent with 
the upper limits for the cold dark matter density. All shaded regions represent excluded parameter space by the corresponding observations.}
\label{fig:cons_frac}
\end{figure}

The PBH mass loss rate slows down according to 
$dM^{'}_{\rm PBH}/dt=S(M_{\rm PBH})^{-k'}dM_{\rm PBH}/dt$ with the value of the exponent $k'>0$, where $dM_{\rm PBH}/dt$ is the standard Hawking rate 
and $S(M_{\rm PBH})$ is the PBH entropy. The lifetime of a memory-burdened PBH is then $\tau \sim RS^{k'}_{\rm BH}$ with $k'=2$ 
being the most conservative estimate. 
Consequently, light PBHs with $M_{\rm PBH}<10^{15}~\rm g$ can survive to the present and continue to emit high‑energy particles 
such as neutrinos and photons~\cite{Chianese:2024rsn,Thoss:2024hsr}. 
Furthermore, two memory‑burdened PBHs may merge to form a new black hole with a mass roughly equal to the sum of the progenitor masses. The newly formed black hole can then emit high‑energy particles via Hawking radiation~\cite{Zantedeschi:2024ram}. Using observations of high‑energy neutrinos 
(all flavor neutrino limits from, e.g., the IceCube EHE upper bound) and gamma‑rays (including data from HERO, INTEGRAL, COMPTEL, EGRET, and Fermi‑LAT for both Galactic and extragalactic gamma‑ray fluxes), the authors of Refs.~\cite{Zantedeschi:2024ram,Thoss:2024hsr,Chianese:2024rsn} derived constraints on the current fraction $f_{\rm PBH}$ of such light PBHs. These limits are obtained by requiring that the predicted fluxes of high‑energy photons and neutrinos from memory‑burdened PBHs do not exceed the observed levels. The resulting constraints are summarized in Fig.~\ref{fig:cons_frac}. The blue shaded region shows limits derived from gamma‑ray observations, the red shaded region corresponds to limits from neutrino observations, and the black shaded region represents limits obtained from neutrinos produced by merged PBHs.

We note that Eq.~(\ref{eq:initial_f}), originally derived for standard Hawking evaporation, can also be applied to memory-burdened primordial black holes (PBHs) to derive constraints, despite their non-standard radiation. The key effect of memory burden is that after a PBH loses half of its initial mass via standard Hawking radiation, the evaporation slows down dramatically, allowing light PBHs to survive to the present day. Under the usual assumption that the memory burden activates exactly at this threshold (the most conservative estimate, see~\cite{Dvali:2024hsb}), and that the post-activation lifetime far exceeds the age of the Universe for our mass range ($\sim 10^{3}-10^{9}~\rm g$), the present-day mass is $M_{\rm cur}\simeq M_{i}/2$. 

Assuming comoving number density conservation during the standard evaporation phase (the number of PBHs remains constant, only the mass decreases), we can rescale Eq.~(\ref{eq:initial_f}) to relate the initial mass fraction $\beta(M_i)$ to the current abundance $f_{\rm PBH}(M_{\rm cur})$:

\beqa
\beta(M_i) = && 1.6\times10^{-25} f_{\mathrm{PBH}}(M_{\mathrm{cur}}/\mathrm{g})^{1/2} 
\times \left(\frac{M_i}{M_{\mathrm{cur}}}\right)^{1/2} \nonumber \\ 
= && 1.6\times10^{-25} f_{\mathrm{PBH}} (2M_{\mathrm{cur}}/\mathrm{g})^{1/2},
\label{eq:beta}
\eeqa
where we have used $M_{i}=2M_{\rm cur}$. Neglecting any further evaporation after activation introduces a negligible impact on the final 
$\mathcal{P}_\mathcal{R}$ constraints compared to other theoretical uncertainties (e.g., the activation threshold and the exponent $k'$ 
in the lifetime scaling $\tau\sim RS^{2}_{\rm BH}$). Moreover, this assumption is conservative, as it yields weaker upper limits.


\subsection{Constraints on the primordial curvature power spectrum at small scales}

The new constraints on the (initial) mass fraction of light PBHs are displayed in Fig.~\ref{fig:cons_frac}. For comparison, we also include the two main existing limits from earlier studies: one derived from the production of stable Planck-mass relics remaining after PBH evaporation, and the other from the lightest supersymmetric particles generated via PBH evaporation, both of which must not exceed the observed cold dark matter density. 
These limits are labeled “DM relic density” and “LSP” in Fig.~\ref{fig:cons_frac} (right $y$-axis)~\cite{Josan:2009}.

Using these updated bounds on the abundance of light PBHs together with the relations summarized in Sec.~\ref{sec:mass_fraction}, we derive upper limits on the amplitude of primordial curvature perturbations on very small scales. 
When converting PBH abundance limits into constraints on $\mathcal{P}_\mathcal{R}$, we assume that the power spectrum is 
scale-independent over the narrow range of scales relevant to each PBH mass. 
Specifically, using the relation between $\beta(M_{\mathrm{PBH}})$ and $\sigma(R)$ from Eq.~(\ref{eq:betampbh}), together with the expression for $\sigma^2(R)$ in Eq.~(\ref{eq:sigma2}), we constrain the primordial power spectrum by assuming that it is locally scale-invariant at each scale $k$, since the integral in Eq.~(\ref{eq:sigma2}) is dominated by modes with $k \sim R^{-1}$. Hence, for each PBH mass $M_{\mathrm{PBH}}$ corresponding to a smoothing scale $R$, the amplitude $\mathcal{P}_{\mathcal{R}}(k)$ is effectively proportional to $\sigma^2(R)$~\cite{PhysRevD.100.063521}, and we can directly translate the abundance limits into constraints on $\mathcal{P}_{\mathcal{R}}$. For a given PBH mass $M_{\rm PBH}$, 
the corresponding comoving wavenumber is $k=k(M_{\rm pbh})$. We then treat $\mathcal{P}_\mathcal{R}$ as a constant over a logarithmic interval 
around that $k$ and compute the mass variance $\sigma^{2}(R)$ using Eq.~(\ref{eq:sigma2}). The derived upper limit on $\mathcal{P}_\mathcal{R}$ 
for that $k$ is thus the value that would produce the observed bound on $\beta$ of $f_{\rm PBH}$. 
This approach is standard in the literature (see, e.g., Refs.~\cite{Josan:2009,Bringmann:2011ut}). 
The comoving wavenumber $k$ corresponding to a PBH of mass $M_{\rm PBH}$ is given by the horizon scale at formation. 
During radiation domination, the relation is~\cite{Josan:2009}\footnote{The mass of a PBH formed in the early Universe is $M_{\rm PBH}=\gamma M_{\rm H}$, 
where $\gamma=0.2$ is the typical value denoting the fraction of the horizon mass that collapses into a PBH. 
$M_{\rm H}$ is the horizon mass at horizon entry, and it is related to the smoothing scale $R$ as 
$M_{\rm H}=M^{\rm eq}_{\rm H}(k_{\rm eq}R)^{2}(g^{\rm eq}_{\star}/g^{i}_{\star})^{1/3}$, where $g^{\rm eq,i}_{\star}$ 
are the total numbers of effectively massless degrees of freedom at matter-radiation equality and at the PBH formation time, respectively. 
We set $g^{\rm eq}_{\star}=3$ and $g^{i}_{\star}=100$. $M^{\rm eq}_{\rm H}$ and $k_{\rm eq}$ are the horizon mass 
and comoving wavenumber at matter-radiation equality, respectively.  
Using the relation $k=2\pi/R$, one then obtains Eq.~(\ref{eq:kmpbh}). }

\beqa
k = 1.4 \times 10^{23} {\rm Mpc^{-1}} \left(M_{\rm PBH}/{\rm g}\right)^{-1/2}.
\label{eq:kmpbh}
\eeqa 

\begin{figure}
\centering
\includegraphics[width=0.5\textwidth]{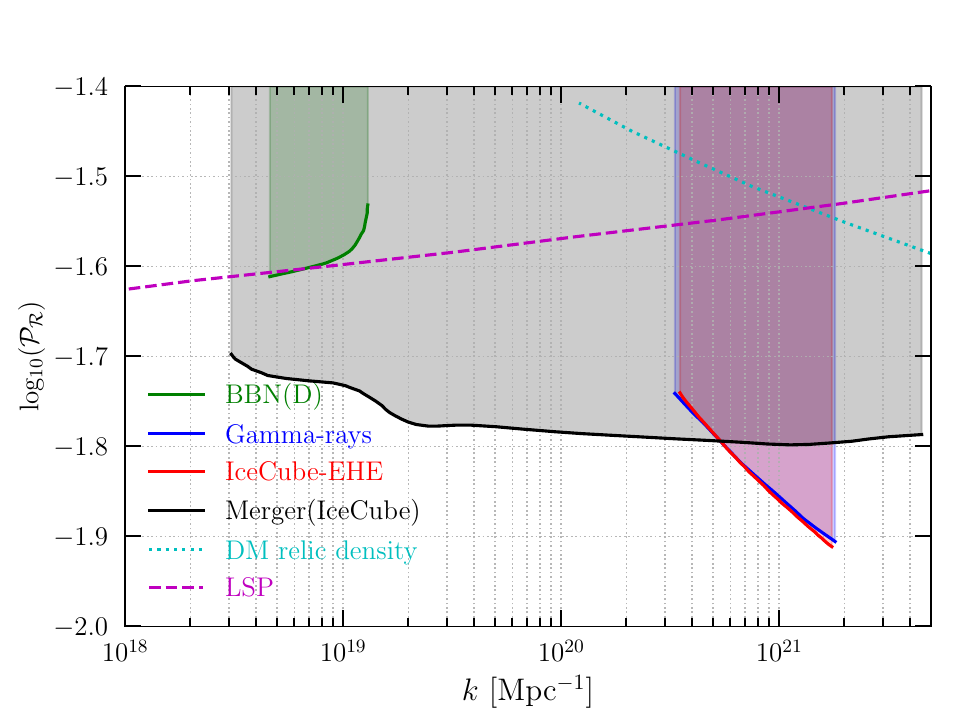}
\caption{Constraints on the power spectrum of the primordial curvature perturbation ($\mathcal{P}_\mathcal{R}$) for small scales 
$3\times 10^{18}~{\rm Mpc^{-1}}\lesssim k \lesssim 4.5\times 10^{21}~\rm Mpc^{-1}$. Line styles follow those in Fig.~\ref{fig:cons_frac}. 
For comparison, previous constraints are also shown: 
(i) a stable Planck-mass relic from primordial black hole evaporation, 
consistent with the upper limits for the cold dark matter density (``DM relic density'', dotted line); 
(ii) the lightest supersymmetric particles generated by PBH evaporation(``LSP'', dashed line), 
consistent with the upper limits for the cold dark matter density~\cite{Josan:2009}.
}
\label{fig:cons_pps}
\end{figure}

The resulting constraints on 
$\mathcal{P}_\mathcal{R}$ are presented in Fig.~\ref{fig:cons_pps}, covering the scale range 
$3\times 10^{18}\lesssim k\lesssim 4.5\times 10^{21}~\rm Mpc^{-1}$, a regime that has been underexplored in previous studies. 
These constraints are derived using the following procedure. For each PBH constraint curve 
in Fig.~\ref{fig:cons_frac} and for each PBH mass, we first compute the corresponding wavenumber $k$ via Eq.~(\ref{eq:kmpbh}). 
We then combine Eqs.~(\ref{eq:sigma2}), (\ref{eq:betampbh}), and (\ref{eq:p}) to solve for $\mathcal{P}_\mathcal{R}$. More specifically, 
given $\beta(M_{\rm PBH})$ or $f_{\rm PBH}$, which is converted to $\beta(M_{\rm PBH})$ using Eq.~(\ref{eq:initial_f}), 
we use Eq.~(\ref{eq:betampbh}) to obtain $\sigma(R)$, adopting $\delta_{c}=0.42$. Finally, assuming a nearly scale-invariant $\mathcal{P}_\mathcal{R}$ 
on scales around $R\sim 1/k$, we infer $\mathcal{P}_\mathcal{R}$ from the corresponding value of $\sigma(R)$.

For comparison, we also show the earlier limits on $\mathcal{P}_\mathcal{R}$ obtained from the $\beta_{\rm PBH}$ 
constraints displayed in Fig.~\ref{fig:cons_frac}. 
On scales $3\times 10^{18}\lesssim k\lesssim 5\times 10^{20}~\rm Mpc^{-1}$, the tightest bounds come from high-energy neutrinos emitted by merged memory‑burdened PBHs, giving $\mathcal{P}_\mathcal{R}\lesssim 10^{-1.8}$. In the scale range $5\times 10^{20}\lesssim k\lesssim 1.8\times 10^{21}~\rm Mpc^{-1}$, the strongest limits are set by non‑merged memory‑burdened PBHs, with neutrino-based constraints being comparable to those from gamma‑rays. BBN‑derived constraints, corresponding to scales $7\times 10^{18}\lesssim k\lesssim 1.3\times 10^{19}~\rm Mpc^{-1}$, are weaker than those from memory-burdened PBHs. However, BBN still provides robust limits on the complementary range 
$4.5\times 10^{18}\lesssim k\lesssim 7\times 10^{18}~\rm Mpc^{-1}$. Overall, the new constraints on $\mathcal{P}_\mathcal{R}$ obtained from memory‑burdened PBHs are significantly tighter than the previous limits derived from dark matter relics or 
supersymmetric particles (“DM relic density” and “LSP”). 

Note that although we have employed previous limits on PBHs and adopted a critical density threshold of $\delta_c = 0.42$ in our calculations, these results are consistent with each other. In fact, the earlier PBH constraints we used were derived without specifying the detailed formation criteria of PBHs.~\footnote{One related issue is the assumption of a monochromatic PBH mass distribution adopted in this work, which is tied to the formation criterion, although other mass functions have also been proposed in different formation scenarios~\cite{Carr:2020gox}.} For our purposes, the formation criteria become important and are only required after obtaining the limits on $f_{\rm PBH}$ and converting them into limits on $\mathcal{P}_\mathcal{R}$.


\section{Summary}

Constraints on the primordial curvature perturbation power spectrum at small scales are largely derived from studies of primordial black holes (PBHs). In particular, previous work on scales $10^{17}\lesssim k\lesssim 10^{23}~{\rm Mpc^{-1}}$ has set limits on $\mathcal{P}_\mathcal{R}$ 
based on the abundance of the lightest supersymmetric particles produced by PBH evaporation and of stable Planck‑mass relics that may remain after evaporation. 

Light PBHs with mass $M_{\rm PBH}\lesssim 10^{9}~\rm g$ were long thought to evade big bang nucleosynthesis (BBN) constraints because their evaporation timescale ($\tau \lesssim 0.01~s$)is much shorter than the onset of BBN. However, recent studies have shown that even such light PBHs, in the mass range $10^{8}\lesssim M_{\rm PBH}\lesssim 10^{9}~\rm g$, can alter the expansion rate of the Universe and the baryon‑to‑photon ratio, thereby modifying the predicted primordial abundances of light nuclei. Consequently, observed light‑element abundances can be used to place limits on the initial mass fraction of these PBHs. 

While standard Hawking evaporation predicts that PBHs with $M_{\rm PBH}\lesssim 10^{15}\rm g$ should have completely evaporated by now, recent theoretical developments suggest that a “memory burden” effect, which was neglected in the original Hawking picture, 
becomes important, e.g., once a black hole loses roughly half its mass. This effect substantially slows down the mass‑loss rate, allowing light PBHs to survive until today and continue emitting high‑energy particles. Furthermore, two such memory‑burdened PBHs may merge to form a new black hole of roughly the combined mass, which subsequently radiates via Hawking emission. Observations of high‑energy neutrinos and gamma‑rays can therefore be used to constrain the current abundance of these light PBHs, which in turn probes the primordial curvature perturbations on correspondingly small scales.

Building on recent advances in black hole physics and updated limits for the (initial) mass fraction of light PBHs in the range 
$10^{3}\lesssim M_{\rm PBH}\lesssim 2\times 10^{9}~\rm g$, 
we have derived new constraints on the primordial curvature perturbation power spectrum on small scales 
$3\times 10^{18}\lesssim k\lesssim 4.5\times 10^{21}~\rm Mpc^{-1}$, which have rarely been studied in previous literature. 
The strongest limits come from the memory-burdened PBHs scenario. 
Note that this work does not derive new PBH abundance limits from raw data. Instead, it takes the latest constraints on light PBHs from BBN 
and from memory‑burdened PBH phenomenology and, for the first time, translates them into upper bounds on $\mathcal{P}_\mathcal{R}$ 
across the uncharted scale range. The resulting limits are substantially stronger and extend to much smaller scales than previous 
relic‑based constraints.

It is important to note that our new constraints on $\mathcal{P}_\mathcal{R}$ from memory-burdened PBHs are model-dependent. 
They rely on the assumption that the memory burden sets in exactly at $M = M_{i}/2$ and that subsequent mass loss is negligible. 
Theoretical uncertainties in the value of the exponent $k'$ and the activation threshold could shift the limits. 
Nevertheless, under this well-motivated scenario, the resulting bounds are significantly stronger than previous ones derived from stable relics and LSPs. 
Future high‑energy neutrino detectors such as IceCube‑Gen2 and GRAND200k are expected to provide even tighter constraints 
in this range~\cite{Chianese:2024rsn,IceCube:2019pna,IceCube-Gen2:2020qha,GRAND:2018iaj}.

\section{Acknowledgements}
This work is supported by the Shandong Provincial Natural Science Foundation (Grant No.ZR2025MS16).

\bibliographystyle{apsrev4-1}
\bibliography{refs}

\end{document}